\documentstyle[twocolumn,aps,psfig]{revtex}

\begin{document}
% \draft command makes pacs numbers print
\draft
%\wideabs{
\title{ Thermal Conductivity as a Probe of Quasi-Particles in the Cuprates.}
% repeat the \author\address pair as needed
\author{N. P. Ong, K. Krishana, Y. Zhang and Z. A. Xu$^\dagger$}
\address{ Joseph Henry Laboratories of Physics, 
Princeton University, Princeton, New Jersey 08544, U. S. A.}
\date{\today}
\maketitle
\begin{abstract}
In underdoped ${\rm YBa_2Cu_3O_x} \;\;(x=6.63)$, the low-$T$ thermal 
conductivity $\kappa_{xx}$ varies steeply with field $B$ at small $B$, 
and saturates to a nearly field-independent value at high fields.  
The simple expression $[1+p(T)|B| ]^{-1}$ 
provides an excellent fit to $\kappa_{xx}(B)$ over a wide 
range of fields.  From the fit, we extract the zero-field 
mean-free-path, and the low temperature behavior of the $QP$ current
The procedure also allows the $QP$ Hall angle $\theta_{QP}$ to be obtained.  
We find that $\theta_{QP}$ falls on the $1/T^2$ curve extrapolated from
the electrical Hall angle above $T_c$.  Moreover, it shares
the same $T$ dependence as the field scale $p(T)$ extracted from
$\kappa_{xx}$.  We discuss implications of these results.
\end{abstract}
% insert suggested PACS numbers in braces on next line
%\pacs{}
%}
% body of paper here
\vskip3mm\noindent
{\bf 1.  Introduction}
\vskip3mm\noindent
Thermal conductivity is potentially a very useful probe of the quasi-
particle excitations in the superconducting state of the cuprates because it 
is capable of detecting the quasi-particle ($QP$) current in the bulk 
\cite{Hagen,Yu,Richardson,Krish1}.
In addition, measurements of its field dependence may yield quantitative
information on the $QP$ mean free path.  At present, this seems to be the
best way to investigate the low-lying excitations of the condensate.
However, the task of disentangling the $QP$ current from the larger 
phonon current in the cuprates poses a difficult problem for experiment.  

We report recent experiments in which the 
direct separation of the $QP$ current is achieved by detailed
analysis of the field dependence of the longitudinal conductivity $\kappa_{xx}$.
This line of approach was motivated by the observation of plateau 
features in high-purity $\rm Bi_2Sr_2CaCu_2O_8$ (Bi 2212) \cite{Krish2}.  
The existence of the plateaus at low $T$ (where $\partial 
\kappa_{xx}/\partial H = 0$) implies that, in the cuprates, vortices are essentially 
transparent to the phonons.  Extensions of these 
measurements to underdoped $\rm YBa_2Cu_3O_x$ (YBCO) reveal 
that this result may be a rather general feature 
of the cuprates in the clean-limit.  This seems to us a significant finding 
since it allows a direct separation of the $QP$ current by the application of 
an intense field.  In addition, we find that the zero-field mean-free-path 
$\ell_0$ of the $QP$ may be estimated to within a factor equal to the 
vortex scattering cross-section $\sigma_{tr}$. 

The isolation of the $QP$ current allows more specific information to be
extracted from the thermal Hall conductivity $\kappa_{xy}$ (Righi-Leduc 
effect)\cite{Krish1}.  With the electronic current independently determined,
we may now obtain the Hall angle $\tan \theta_{QP}$.  The $QP$ Hall angle
uncovers a number of interesting features which we discuss below.

\vskip3mm\noindent
{\bf 2.  Experiment}
\vskip3mm\noindent
Measurements of $\kappa_{xx}$ in the mixed state of the cuprates have been 
reported by several groups.
\cite{Richardson,Krish1,Krish2,Uher,Ong,Aubin}
Experiments in intense magnetic fields $\bf H$ are complicated by problems such as 
cleaving of the crystal by the large torques generated.  The most serious problem, 
however, seems to stem from the field sensitivity of the thermometers.  At 
the resolution needed, the field dependence of the sensors is serious (in 
thermocouples, moreover, the field sensitivity is also history dependent).
Previously, we employed a bridge-balance method to get around the field-
sensitivity problem \cite{Krish2}.  
\begin{figure}[h]
\centerline{\psfig{figure=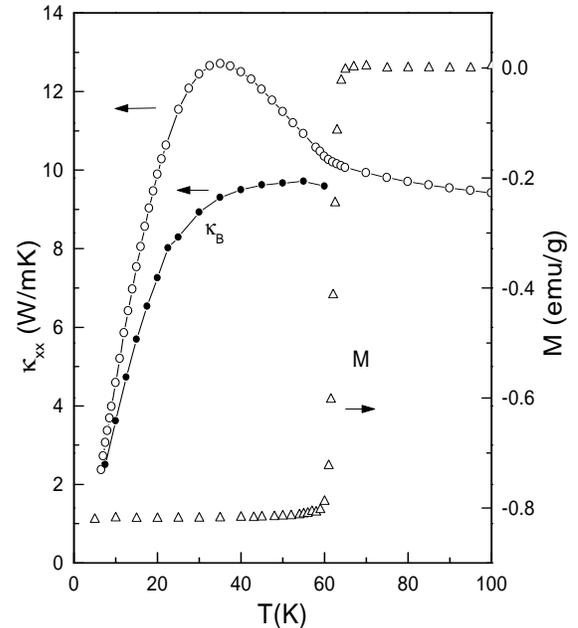,height=3.3in,width=3.1in}}
\caption{ The $T$ dependence of the in-plane thermal conductivity 
$\kappa_{xx}(0)$ in zero field (open symbols) in underdoped YBCO with $x$ = 
6.63 and $T_c$ = 63 K.  The solid symbols are the phonon conductivity 
$\kappa_B$ estimated by the fit of the thermal 
magnetoconductance to Eq. \ref{kvsB}. The $dc$ magnetization of the crystal is shown as open %%@
triangles ($H$ = 5 Oe).}
\label{F1}
\end{figure} 
\noindent
In our present approach, we adopt a single-heater, 
two-sensor method in which the temperature difference $\delta T $ between the 
ends of the sample is detected by two closely matched resistive sensors 
(cernox).  The thermal gradient $-\nabla T$ is applied in the $ab$ plane, 
and $\bf H$ is parallel to $\bf c$. The temperature is regulated 
(with a third base-cernox), and measurements are taken after waiting 
about 10 min. for the field to stabilize.  The readings of the 
two sensors are recorded at three values of the heater current 
$I$ = 0, 0.4, 0.6 mA (typically).  The $I$ = 0 readings are used to 
calibrate the field dependence of the two cernox sensors, while the values 
of $\delta T$ determined with $I$ at the two values provide a 
check on the linearity of the sample response.  By 
testing with a standard material that has no 
intrinsic field dependence in $\kappa_{xx}$ (nylon), we have found 
that at temperatures above 8 K this method provides an accurate and 
highly reproducible determination of the intrinsic conductance to a 
resolution of 1 in $10^3$.  Although the bridge-balance method is capable of 
higher resolution, the present technique lends itself to full automation.  A 
higher density of points may be obtained, and checks (e.g. for linearity) 
can be made {\em in situ}.  A pair of thermocouple junctions are used to detect
the $H$-antisymmetric (Hall) gradient to obtain $\kappa_{xy}$.

With the high sensitivity, we have found that, in untwinned, 
optimally-doped $\rm YBa_2Cu_3O_x$ ($T_c$ = 93 K, $x$ = 6.95), 
$\kappa_{xx}$ in a field $\bf H \parallel c$ 
becomes increasingly hysteretic below 35 K.  Although the hysteresis is 
small (about 5$\%$ of the total $\kappa_{xx}$ at 8 K), it greatly 
complicates the 
extraction of the $QP$ current (we discuss this later).  In underdoped 
crystals, however, the hysteresis is unobservable up to 14 
tesla (less than $10^{-3}$), and the observed $\kappa_{xx}$ vs. $H$ 
is a faithful representation of its intrinsic dependence on $B$.  

In this report, we discuss data from a twinned, underdoped crystal 
in which $T_c$ = 63 K, and $x$ = 6.63.  The zero-
field temperature profile of $\kappa_{xx}$ is shown in 
Fig. \ref{F1}.  The relative magnitude of the anomaly in 
$\kappa_{xx}$ is only about a quarter of that in the 93-K YBCO, but 
larger than in optimum Bi 2212 \cite{Krish2} and 
$\rm La_{2-x}Sr_xCuO_4$ (LSCO) \cite{Ong}.  Also 
shown (solid symbols) is our new estimate of the phonon conductivity 
($\kappa_B$).  One of our main results is that the entire 
field dependence of $\kappa_{xx}$ derives from the $QP$ 
current, while the phonon current is unaffected by $H$.

\begin{figure}[h]
\centerline{\psfig{figure=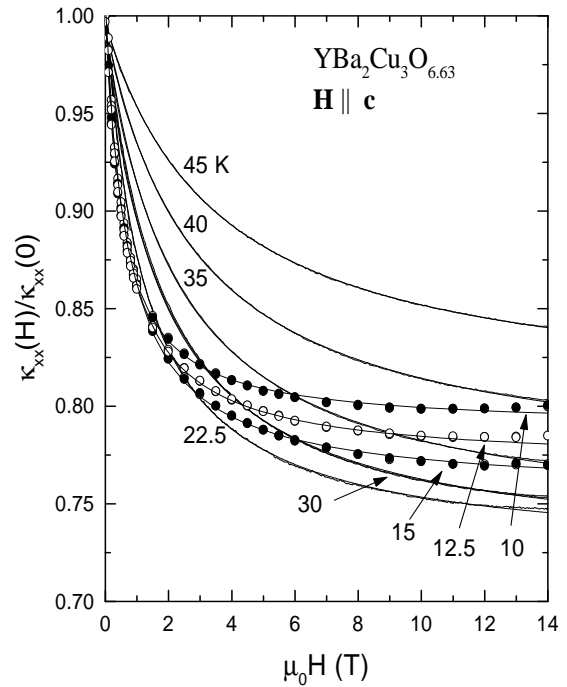,height=3.6in,width=3.1in}}
\caption{Variation of $\kappa_{xx}$ with field $\bf H \parallel c$ in underdoped $\rm %%@
YBa_2Cu_3O_{6.63}$ at indicated temperatures.  All curves are non-hysteretic to 1 part in $10^3$ %%@
(the curves above 15 K are superpositions of sweep-up and sweep-down traces).  A fit to the data %%@
at 22.5 K is also shown superposed.  Curves at 15 K and below (discrete symbols) show the %%@
approach of $\kappa_{xx}$ to the $H$-independent value ($\kappa_B$).}
\label{F2}
\end{figure}

\begin{figure}[h]
\centerline{\psfig{figure=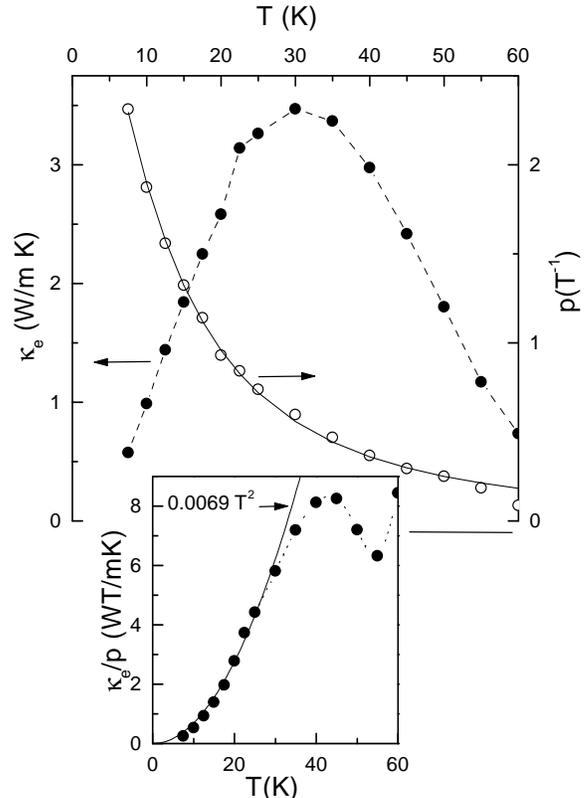,height=4.2in,width=3in}}
\caption{ The $T$ dependence of the zero-field $QP$ conductivity $\kappa_e$ and the field scale %%@
$1/p(T)$ extracted from fits to $\kappa_{xx}$ vs. $H$ in underdoped YBCO. The $T^2$ dependence of %%@
$1/p(T)$ suggests that it is related to a scattering rate. The solid line is $c_0+c_1T + c_2T^2$, %%@
with $c_0$ = 0.25, $c_1$=0.016, and $c_2$= 0.0012 (SI units). The inset plots the $T$ dependence %%@
of the ratio $L_e(T)\equiv \kappa_e(T)/p(T)$ (in which $\ell_0$ cancels).  In the vortex %%@
scattering model, $L_e$ should vary as $T^2$ in a $d$-wave superconductor at low $T$ (Eq. %%@
\ref{ke})}
\label{F3}
\end{figure}

\vskip3mm\noindent
{\bf 3. Results and Analysis}
\vskip3mm\noindent
Figure \ref{F2} displays the field dependence of $\kappa_{xx}$ 
at selected $T$.  With decreasing $T$, the initial slope 
of $\kappa_{xx}$ increases rapidly.  Below 15 K, the 
rapid decrease crosses over to an almost flat dependence, which recalls the 
plateau features observed in single-domain Bi 2212 \cite{Krish2}.  Within our resolution, there %%@
is no resolvable hystereses at the temperatures investigated.  The 
higher precision and larger range of the new data enable us to compare the 
observed field dependence with various expressions.  We find that the field 
dependence is accurately fitted (Fig. \ref{F2}) to the expression 
\begin{equation}
\kappa_{xx}(B,T)= \frac{\kappa_e(T)}{(1+p(T)|B|)} + \kappa_B(T),
\label{kvsB}
\end{equation}
where the entire $B$ dependence resides in the denominator of the first term, 
and the term $\kappa_B$ is a field-independent background.  
At each temperature, the fit yields the 
two parameters $\kappa_e(T)$ and $p(T)$ associated with the 
$QP$ current, and the term $\kappa_B$ which we identify with 
the phonon term, viz. $\kappa_B$ = $\kappa_{ph}$.  
\noindent
In our previous experiments on optimally-doped YBCO 
\cite{Krish1} and LSCO \cite{Ong}, fits to Eq. \ref{kvsB}
were ambiguous because of strong hystereses (the distortions introduced 
caused the extracted $p(T)$ to be non-monotonic in $T$).  These extraneous 
effects led us to consider alternate expressions (see below), as well as a 
possibly field dependent $\kappa_{ph}$.  However, the present results 
have clarified the problem.  In addition to the absence of 
observable hysteresis, the curves for $\kappa_{xx}$ at low $T$ 
display pronounced curvature in moderate fields, 
corresponding to a strong attenuation of the $QP$ current.  The attenuation 
uncovers a nearly field-independent background that we identify with the 
phonon thermal conductivity.  In the main panel of Fig. \ref{F3}, 
we display the $T$ depenence of the parameters $\kappa_e$ and $p(T)$. 

The motivation for Eq. \ref{kvsB} is that, near the vortex core, the steep 
variation of the pair potential and the circulating superfluid together 
present a strong scattering potential for an incident $QP$ \cite{Cleary}.  
Expressing the scattering rate as a transport cross-section 
$\sigma_{tr}$, we assume additivity of the 
rates, and write the $QP$ mean-free-path in a 
field as  $\ell(B) = \ell_0/[1+\ell_0\sigma_{tr}|B|/\phi_0]$, 
where $\ell_0$ is the {\em zero-field} value of 
the mean-free-path, and $\phi_0$ the flux quantum.  Thus, in this model, 
the $QP$ conductivity is given by Eq. \ref{kvsB} 
with \cite{Cleary,Vinen,Krish1}
\begin{equation}
p(T) = \ell_0\sigma_{tr}/\phi_0.
\label{p}
\end{equation}

In the Boltzmann equation approach, we may write the $QP$ 
thermal conductivity (in zero field) as \cite{Bardeen}
\begin{equation}
\kappa_e(T) = \frac{1}{T}\sum_{\bf k} (-\frac {\partial f }
{\partial E} ) E({\bf k})^2 v_x({\bf k})^2 \tau({\bf k}),
\label{ke}
\end{equation}
where $E(\bf k)$ and $\tau(\bf k)$ are, respectively, the 
energy and lifetime of a $QP$ in the state $\bf k$, and 
${\bf v(k)} = \hbar^{-1}{\bf \nabla} E(\bf k)$ its group 
velocity.  In a $d$-wave superconductor at low $T$, 
the excitations are confined 
to Dirac cones at the nodes where the energy 
may be parametrized as $E(k_1, k_2) = \hbar\sqrt{ (k_1v_f)^2 + (k_2v_2)^2 }$
($k_1$ and $k_2$ are the components of $\bf k$ normal and parallel 
to the Fermi Surface), $\kappa_e$ reduces to 
\begin{equation}
\kappa_e(T)= \frac{\eta}{\pi}\frac{k_B^3T^2} {\hbar^2}\frac{\ell_0}{v_2}(1+\frac{v_2^2}{v_f^2}),
\label{T2}
\end{equation}
where $\ell_0 = v_f\tau_0$ is the mean free path at the nodes, and $\eta\equiv
\int^\infty_0 dx\;x^3(-df/dx)\;\sim 5.41$.

The $T^2$ dependence of $\kappa_e$ in Eq. \ref{T2} is masked by the strong $T$ 
dependence of $\ell_0$.  However, in our experiment, the latter is obtained 
independently from the field dependence (with $p(T)$ given by Eq. \ref{p}).  
We may divide out the $T$ dependence of $\ell_0$, to isolate 
the quantity $L_e(T) = \kappa_e(T)/p(T)$.  Comparing 
Eqs. \ref{p} and \ref{ke}, we are left with an 
expression that contains only two 
material-specific parameters $v_2$ and $\sigma_{tr}$, viz.
\begin{equation}
L_e(T)= \frac{\eta}{\pi}\frac{k_B^3T^2} {\hbar^2}\frac{\phi_0}{v_2\sigma_{tr}},
\label{Le}
\end{equation}
Figure \ref{F3} (inset) reveals that the measured $L_e(T)$ 
displays a nearly $T^2$ dependence 
at low $T$, that may be fitted to give $L_e(T) = 6.9\times 10^{-3} T^2$ WT/mK.  
Comparing this expression with Eq. \ref{Le}, we determine from our experiment
\begin{equation}
v_2\sigma_{tr} = 2.11 \times 10^{-4} {\rm m^2/s}.
\label{v2}
\end{equation}

\begin{figure}[h]
\centerline{\psfig{figure=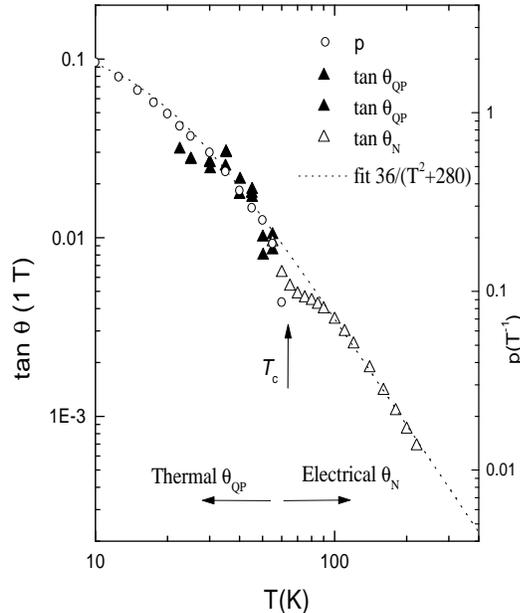,height=3.2in,width=3in}}
\caption{The temperature dependences of the field scale $p(T)$ (open circles) and the weak-field %%@
$QP$ Hall angle below $T_c$ (closed triangles) and the electrical Hall angle $\tan\theta_N$ above %%@
(open triangles).  The close similarity of $T$-dependences of $p$ and $\tan \theta_{QP}$ strongly %%@
suggests that $p$ is proportional to the $QP$ lifetime.  The broken line shows that the $QP$ %%@
weak-field Hall angle is numerically equal to the extrapolation of $\tan\theta_N$.  }
\label{F4}
\end{figure}

\vskip3mm\noindent
{\bf 4. The Hall angle}
\vskip3mm\noindent
The field parameter $p(T)$ has been extracted from $\kappa_{xx}$ alone.  While its nominally %%@
$1/T^2$ variation (Fig. \ref{F3}) is consistent with the identification $p(T)\sim \ell_0$ (Eq. %%@
\ref{p}), it is important to see if this is consistent with a separate experiment. We turn next %%@
to the Hall conductivity $\kappa_{xy}$.  In the previous Hall study on optimal YBCO\cite{Krish1}, %%@
$\kappa_{xy}$ was analyzed without the benefit of information on the diagonal electronic current.  %%@
From the analysis above, we may now extract the Hall angle $\tan\theta_{QP}(H)= %%@
\kappa_{xy}(H)/\kappa_e(H)$ as a continuous function of $H$ at each temperature.  In general, %%@
$\tan\theta$ displays strong negative curvature vs. $H$ \cite{Krish4}.  Here, we restrict our %%@
discussion to the weak-field value $\theta_{QP}(0)$.  In underdoped YBCO, the small $QP$ %%@
population generates a weak thermal Hall current, and the uncertainties in determining %%@
$\theta_{QP}$ are quite large (compared to optimum YBCO)\cite{Krish4}.  Nevertheless, we find two %%@
interesting features of the Hall angle (see Fig. \ref{F4}).  First, $\theta_{QP}(0)$ (solid %%@
triangles) and $p$ (open circles) share the same $T$-dependence from $T_c$ to about 20 K.  %%@
Secondly, we recall that the normal-state electrical Hall angle $\theta_N$ (open triangles) %%@
follows a $1/T^2$ dependence. 

\noindent
 The new values for $\tan\theta_{QP}(0)$ lie on the curve for $\theta_N$ extrapolated below %%@
$T_c$.  As shown in Fig. \ref{F4}, the three quantities $p$, $\tan\theta_{QP}(0)$ and %%@
$\tan\theta_N$ fall on the same curve over about 2.5 decades (in the plot $p$ and %%@
$\tan\theta_{QP}(0)$ are related by a constant scale factor).  Just above $T_c$, $\tan\theta_N$ %%@
displays a slight dip associated with fluctuation effects.  The similarity 
between $\theta_{QP}$ and $\theta_N$ has also been pointed out by Zeini {\em et al.} %%@
\cite{Zeini}.

We briefly discuss our interpretation.  These results suggest that the $1/T^2$ dependence is, in %%@
fact, intrinsic to the $QP$'s below $T_c$.  The ubiquitous $T^2$ dependence of $\cot \theta_N$ in %%@
the normal state appears to be an extension of the low-temperature behavior into the normal %%@
state.  The similarity between the $T$-dependences of $p$ and $\theta_{QP}$ imply that the %%@
diagonal and the Hall channels relax with the same $T$-dependence, $\sim 1/T^2$, consistent with %%@
simple Drude behavior.  Just as in conventional metals, the thermal Hall resistivity in the mixed %%@
state, $W_{xy}\equiv \kappa_{xy}/\kappa_e^2 = \tan\theta_{QP}(0)/\kappa_e$, should provide a %%@
measure of the heat capacity of the $QP$ population, as it does (see inset of Fig. \ref{F3}).  

\noindent
This conventional behavior is abruptly altered when we cross $T_c$ into the normal state.  The %%@
Hall angle continues to relax at the same numerical rate.  By contrast, the scattering rate in %%@
the diagonal conductivity undergoes a dramatic change\cite{Krish4}.  The transport mean-free-path %%@
$\ell_0$ decreases abruptly by a factor of 4-6 (10 in optimal crystals) across $T_c$.  This sharp %%@
decrease, followed by a nominally $T$-linear scattering rate, is responsible for the anomalously %%@
strong $T$-dependence of the Hall coefficient in the normal state.  Thus, the anomalous channel %%@
responsible for most of the strange-metal properties is the diagonal conductivity.  The Hall %%@
channel appears to be quite conventional.  A detailed discussion of these results including %%@
optimal crystals will appear elsewhere\cite{Krish4}.

\vskip3mm\noindent
{\bf 5. Discussion}
\vskip3mm\noindent
We have discussed how the field dependence of $\kappa_{xx}$ in underdoped YBCO 
may be analyzed to extract electronic parameters, such 
as $\ell_0$ and $\sigma_{tr}$.  The 
analysis derives from two striking features of $\kappa_{xx}$ intrinsic 
to high-purity 60-K YBCO crystals at low $T$, namely the steep 
decrease of $\kappa_{xx}$ in weak field followed by a saturation 
at high field, and the absence of resolvable hysteresis.  

The first feature allows us to fit a much broader range of field 
scales (as expressed by the dimensionless parameter $p(T)B$).  The higher 
density of measurements also helps.  We illustrate this point as follows.  In 
a previous attempt \cite{Ong} to analyze similar measurements in LSCO, it was 
found that the $\kappa_{xx}$ vs. $H$ curves were equally well-fitted by 
Eq. \ref{kvsB} and the expression $G(B) = \psi(1/2 + B_0/B) + \ln(B_0/B), $
with $\psi(x)$ the digamma function.  With the data scatter and 
the smaller range of reduced fields 
$B/B_0$ in LSCO, the two fits could not be distinguished, and Ong {\em et al.}
\cite{Ong} argued the case for adopting $G(B)$ to describe $\kappa_{xx}$ 
in LSCO.  However, with the larger field scale $pB$ here, we find that Eq. 
\ref{kvsB} provides a much better fit (this is evident if the 
two fits are compared in a plot versus $\ln H$).  The physics underlying 
the two fit expressions is of course quite different.  In 
light of the present work, we now favor adopting Eq. \ref{kvsB}, 
instead of the digamma function fit, for analyzing $\kappa_{xx}$ 
vs. $H$ curves in cuprates.  

The second feature (no observable hysteresis) relates to the issue 
of remanence and vortex pinning in cuprates.  It is known that the 
relaxation of non-equilibrium flux distributions may produce a slow drift 
in $\kappa_{xx}$ if it is observed a few seconds after 
a change in $H$ is made \cite{Richardson,Uher}.  
Further, vortex pinning effects at low $T$ can lead to step-like 
jumps in $\kappa_{xx}$ when the field sweep direction is changed.  
In optimum YBCO and in strongly overdoped Bi 2212, we find that 
$\kappa_{xx}$ increases step-wise when the sweep direction of 
$H$ is reversed from up to down.  Recently, however, a hysteretic 
loop of the opposite sign has been reported by Aubin et al. \cite{Aubin} in 
Bi 2212 ($\kappa_{xx}$ decreases step-wise when $H$ is swept down).  
At present, the origin of the hystereses in $\kappa_{xx}$ is not 
understood (especially the existence of hystereses with different 
signs).  We note that the magnitude of the hystereses in is 
much smaller than that observed in the magnetization 
$M$ vs. $H$.  In our measurements on underdoped YBCO, no hysteresis 
in $\kappa_{xx}$ is observed for fields as large as 14 T at 
temperatures down to 6 K, even though hystereses are sizeable 
in the $M$ vs. $H$ curves.  In particular, the magnitude of 
$\kappa_{xx}$ at the plateau-like region is not hysteretic.  The absence 
of hysteresis implies that the magnetization is too small to influence the 
measured $\kappa_{xx}$, so we may assume that $B = \mu_0 H$, 
as tacitly assumed in the fits.

By contrast, hysteretic effects cannot be neglected in optimally 
doped YBCO (as discussed above). Stronger vortex pinning is clearly 
responsible for the larger hysteresis in the 93-K crystal.  Below 35 K in 
this crystal, the hysteresis steadily increases to about 5$\%$ 
at 8 K.  Although the hysteresis is small, the remanence produces 
in the trace of $\kappa_{xx}$ vs. $H$ both a broadening at small 
$H$ and an asymmetry about $H$ = 0 that strongly 
distort the fit to Eq. \ref{kvsB}.  The distortions preclude a 
meaningful extraction of below about 35 K.  Thus, of the 
various phases of the cuprates we 
investigated (YBCO, Bi 2212 and LSCO), the underdoped phase of YBCO 
appears to be the most suitable for our purpose of isolating the $QP$ current 
from the total thermal current.  

In high-purity single-domain crystals of Bi 2212, the field 
dependence of $\kappa_{xx}$ displays a distinct break in slope 
in $\kappa_{xx}$ at a characteristic field $H_k$, followed by 
a plateau region in which $\kappa_{xx}$ is nearly 
independent of $H$ \cite{Krish2}.  The field $H_k$, which 
varies approximately as $T^2$, was interpreted as a 
field-induced phase transition, possibly involving a new 
order parameter.  We compare the present results with the two findings in 
Bi 2212, i.e. the kink feature at $H_k$ and the existence of the plateau.  As 
shown in Fig. \ref{F2}, $\kappa_{xx}$ in 60-K YBCO smoothly 
crosses over into the field-independent region at low $T$, 
instead of displaying a sharp kink.  While the 
plateau regime is similar in the two systems, the kink feature signalling a 
phase transition is absent in the YBCO crystal.  A difference between the 
two systems is the electronic anisotropy.  From the resistivity anisotropy 
$\rho_c/\rho_{ab}$ ($\sim 10^5$ compared with $10^3$), Bi 2212 is much 
closer to the $2D$ limit than underdoped YBCO.  Whether 
this is a significant factor is a subject for future investigation. 

In their experiment Aubin {\em et al.}\cite{Aubin} observed 
the value of $\kappa_{xx}$ in Bi 2212 at the plateau 
(at 8 K) to be ~ $1\%$ higher in the field sweep-up 
direction than in sweep-down.  In rough analogy with the magnetization 
profile in the Bean model, they raise the issue that the plateau may be 
associated with, or reflect a specific state of the vortex system.  In our 
response \cite{Krish3}, we pointed out that the hysteresis in 
their sample is about 5 times larger (at 8 K) than in the 
two crystals used by Krishana {\em et al.}\cite{Krish2}.  At higher 
temperatures, 15-20 K, where the plateau is just as prominent, the 
hysteresis is almost unresolved.  The hysteresis is an extrinsic effect 
possibly associated with stronger flux pinning in a more disordered crystal.  
The present results lend a fresh perspective to the question whether the 
plateau is intrinsic to the $QP$ system or the product of a 
particular state of the vortex system.  The YBCO results 
show that, whenever the $QP$ current 
is very strongly suppressed by the available field, the thermal conductivity 
that remains is indeed field-independent and non-hysteretic.  This shows 
that, in YBCO, a plateau regime definitely exists; the lack of observable 
hysteresis shows that it has nothing to do with a particular state of the 
vortex system.  However, to access it, it may be necessary to work in the 
underdoped regime and to use high-purity crystals with weak pinning (as 
discussed above, we are unable to access the plateau region in 90-K 
YBCO).  

In principle, the analysis described yields quantitative information on the quasi-particles.  %%@
Having both the diagonal and off-diagonal conductivities 
available reduces the uncertainties in identifying the measured parameters, as
well as provides consistency checks.  The weak-field Hall angle may be expressed
as a `Hall' mean-free-path $\ell_H \equiv \theta_{QP}\hbar k_F/e$.  The value of $\theta_{QP}$ at %%@
10 K gives $\ell_H\sim 4,200 \AA$.  The similarity of the $T$ dependences
in $\theta_{QP}$ and $p$ implies that $\ell_H$ is proportional to $\ell_0$. 
If the proportionality constant is 1, we may use $\ell_H$ in Eq. \ref{p} to find that %%@
$\sigma_{tr}\sim 90 \AA$, about 1.7 times the diameter of the 
vortex core ($2\xi$) ($\xi\sim 26 \AA$ if the upper 
critical field $H_{c2}\sim$ 50 T).  
Using this value of $\sigma_{tr}$ in Eq. \ref{v2}, we obtain the velocity
$v_2\sim 2.3 \times 10^6$ cm/s.  From the penetration depth variation $\Delta\lambda =
4.3 \; \AA/K$\cite{Hardy}, Wen and Lee \cite{Wen} obtain the velocity anisotropy
$v_f/v_2\sim 7.6$.  With our estimate for $v_2$, we find that the
Fermi velocity $v_f \sim 1.8\times 10^7$ cm/s.  

We acknowledge support from the U.S. Office of Naval Research 
and the U.S. National Science Foundation.   Useful conversations with 
Philip Anderson, Duncan Haldane, Patrick Lee, Louis 
Taillefer and Shin-Ichi Uchida are gratefully acknowledged.
\vskip3mm\noindent
{\em This manuscript will appear in the Proceedings of the Taniguchi Symposium
on the Physics and Chemistry of Transition Metal Oxides 1998, (Springer Verlag 1999).}
\vskip1mm\noindent
$^\dagger${\em Permanent address:  Department of Physics, 
Zhejiang University, Hangzhou, China.}

% figures follow here
%

\end{document}